\documentclass{PoS}

\title{Hunting for the Conformal Window}

\ShortTitle{Hunting for the Conformal Window}

\author{Albert Deuzeman and \speaker{Elisabetta Pallante}\\
        Centre for Theoretical Physics, University of Groningen, 9747 AG, Netherlands\\
        E-mail: \email{a.deuzeman@rug.nl, e.pallante@rug.nl}}

\author{Maria Paola Lombardo\\
        INFN-Laboratori Nazionali di Frascati, I-00044, Frascati (RM), Italy\\
        E-mail: \email{mariapaola.lombardo@lnf.infn.it}}

\abstract{Undeniably, the imminent activity of LHC and the quest for the nature of
 physics beyond the standard model have raised renewed interest in the conformal and 
quasi-conformal behaviour of gauge field theories with matter content. Theoretically 
driven questions seem to now acquire a strong experimental appeal and might guide us towards
 a more 
realistic string theory to field theory connection, originally inspired by the AdS/CFT 
conjecture. In this brief report, we discuss the state of the art of our search for the 
conformal window in the SU(3) colour-gauge theory with fermions in the fundamental 
representation. 
}

\FullConference{The XXVI International Symposium on Lattice Field Theory\\
                 July 14-19 2008\\
                 Williamsburg, Virginia, USA}

\begin{document}

\section{Introduction}

In the early eighties our understanding of the perturbative behaviour of non abelian gauge
 field theories was enriched by works of Caswell \cite{Caswell}, Banks and Zaks 
\cite{Banks:1981nn}. 
It was noticed that a second zero of the two-loop beta function of an SU(3) gauge theory with
$N_f$ massless fermions in the fundamental representation appears before the loss of 
asymptotic
 freedom at $N_f = 16\frac{1}{2}$. This fact implies, at least perturbatively, the
 appearance of an infrared fixed point (IRFP) for the theory. The closer is the number of 
flavours to $N_f=16$, the closer to zero is the coupling at the IRFP. What is the minimal
 flavour content that suffices for the appearance of the infrared fixed point? The answer
 to this question will depend on the nature of the fermion representation, fundamental or 
adjoint or other. With fermions in the fundamental representation the critical number of
 flavours at which the second zero of the beta function occurs is $N_f = 8.05$.      
The dynamics of chiral symmetry breaking and its restoration can significantly modify 
\cite{Miransky} the 
actual value of the critical number of flavours at which the IRFP arises in the theory. 
At the IRFP the theory is conformal and quasi-conformal in its surroundings.

This promptly suggests a plausible theoretical ground to the quasi-conformal behaviour of
walking technicolor theories, and raises intriguing connections with the strongly 
interacting  dynamics which is naturally embedded in some extra dimensional 
brane-world scenarios. 

\begin{figure}
\begin{center}
\includegraphics[width=.5\textwidth]{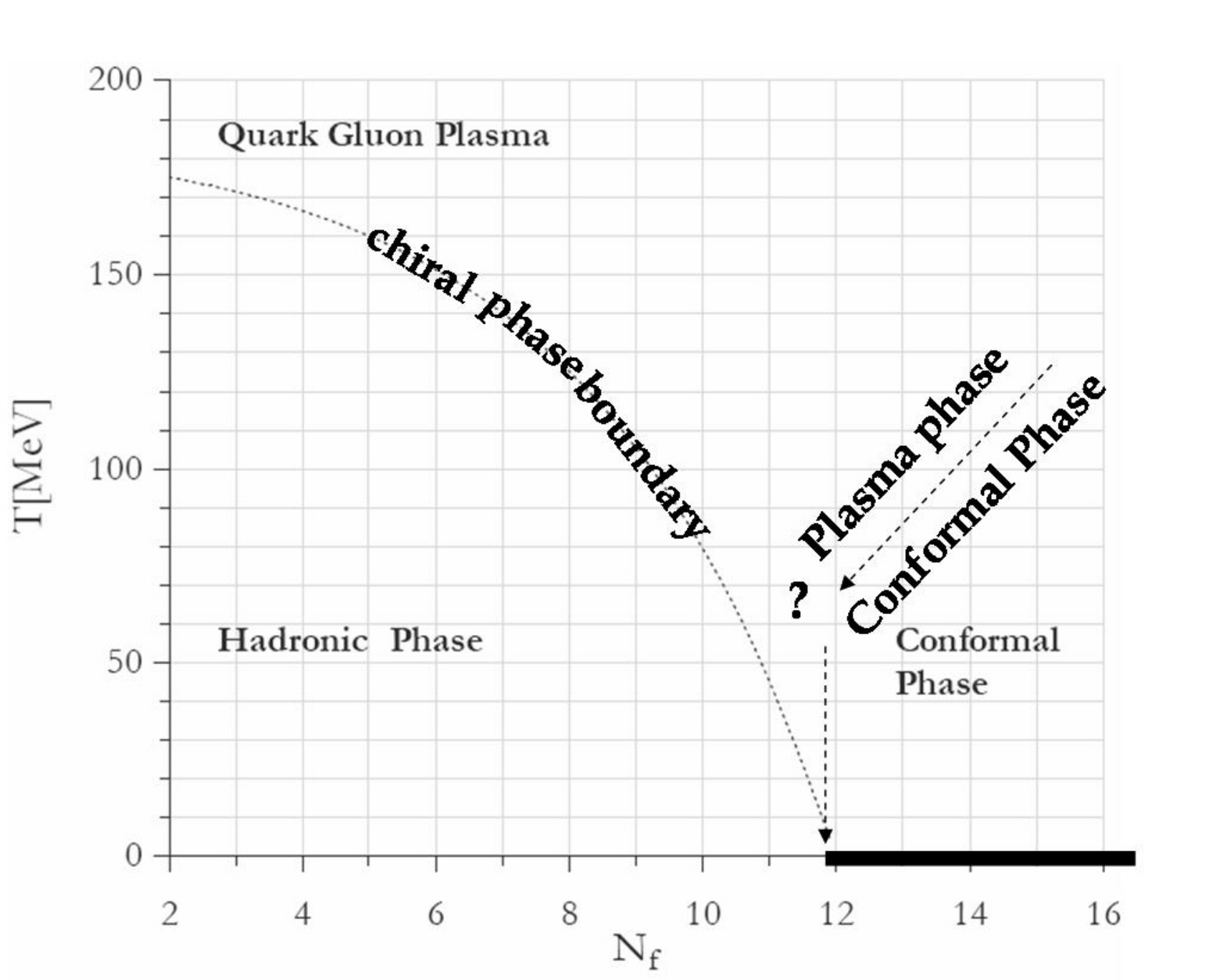}
\caption{A qualitative view of the phase diagram in the $T, N_f$ plane. The lines separating
 the conformal phase from the QGP phase are purely hypothetical and drawn for explanatory 
purposes. The end-point at around $N_f=12$ is suggested by some recent studies 
\cite{Appelquist:2007hu}, \cite{Braun:2006jd}.}
\label{Phaseplot}
\end{center}
\end{figure}

There are at least three reasons why the search for conformality in gauge theories with 
fermionic matter content should be pursued. First as we just said, it offers a theoretical 
framework to walking technicolor-type theories, and more generally to strongly interacting 
type dynamics of new physics at the electroweak symmetry breaking scale.
Second, how is the conformal phase connected to the quark-gluon plasma phase? 
See the sketchy plot in Figure~(\ref{Phaseplot}). Can we better describe the quark-gluon 
plasma characteristics and its deviations from conformality? Third, can we further exploit 
the AdS/CFT conjecture by knowing what mechanisms induce a restoration of the conformal
 behaviour in gauge field theories? 

The critical value of $N_f$ which signals the lower end point of the {\it conformal window}
can well be significantly lower than $N_f=16$, rendering a perturbative study rather 
uncertain. A non perturbative analysis is mandatory. Pioneering lattice studies 
gave conflicting hints towards the critical value of $N_f$ where conformality is restored.
For previous work relevant to this study see \cite{Fukugita:1988}, \cite{Brown:1992fz}, 
\cite{Iwasaki:2003de}, \cite{Damgaard:1997ut}.

In a recent paper \cite{PLB_8flavours} (see also A. Deuzeman, these proceedings) 
we have produced clear indication that eight 
fundamental fermions are still in the QCD-like phase, undergoing a thermal transition to 
the chirally symmetric phase. A recent lattice study of the running coupling 
\cite{Appelquist:2007hu}  suggests that $N_f=12$ should already be in the conformal window.

Our contribution to this enterprise was inspired by a nice plot in \cite{Braun:2006jd}, 
where the shape of the chiral phase boundary is predicted on the base of a truncated 
renormalization group flow approach; the drawback of this framework is the impossibility
to provide an estimate of the size of disregarded contributions after truncation. 
Alternatively, lattice simulations are tailored to this task.
Note that the number of flavours can indeed be regarded as a continuous variable, since
 the Casimir's are polynomials in $N_f$ \cite{Damgaard:1997ut}. This means that the critical
 line found in \cite{Braun:2006jd} is in fact a true phase boundary between a conventional
 Goldstone phase and a chirally symmetric phase, the zero temperature limit of which is the
 onset of the conformal phase.

The scope of our lattice study is not only to locate the conformal window, but to build a 
better understanding of the chiral phase boundary bridging the hadronic phase to the 
quark-gluon plasma phase, and try to explore the region which moves in 
Figure~(\ref{Phaseplot}) from the 
conformal window at zero temperature to the quark-gluon plasma phase at lower number of 
flavours and high temperatures. 

Most tasks in this study involve delicate issues and unexplored aspects of lattice field 
theory. The complete answer to the question if the conformal window exists and where is 
located 
in parameter space can only be formulated by merging all possible complementary lattice 
strategies under a pertubative guidance.

\section{Our strategy}

One possible way of revealing the emergence of an infrared fixed point through a lattice 
simulation of an asymptotically free gauge theory is to observe a flattening in the running 
coupling, as much as it is done in perturbation theory. The framework is the one of the 
Schroedinger functional, and the step-scaling function, and first interesting results have 
been reported in \cite{Appelquist:2007hu} (see also E. Neil, these proceedings). 
In this approach however, more than in others, 
the reduction of systematic uncertainties is essential to discriminate between two different 
scenarios -- the presence or absence of an IRFP -- and prove that the continuum behaviour 
of the running coupling has been reproduced. 
Other interesting proposals study different fermion representations \cite{DeGrand}, or look 
at the eigenvalue spectrum of the Dirac operator \cite{Hollands}.
 
An alternative and complementary strategy, the one we adopted, is inspired by the physics 
of phase transitions and allows for the exploration of multiple aspects of the theory 
in different regimes, e.g. the connection to the quark-gluon plasma phase. 
We look for phase transitions, plotting the 
associated observables at zero and non zero temperature, and we look for critical exponents, 
more specifically the anomalous dimensions of correlation functions in the vicinity of 
fixed points of the theory. The possibility of probing anomalous dimensions of correlation
 functions at the IRFP has also been recently considered in \cite{Luty}. See also 
\cite{Sannino} and refs. therein.

\begin{figure}
\begin{center}
\includegraphics[width=.4\textwidth]{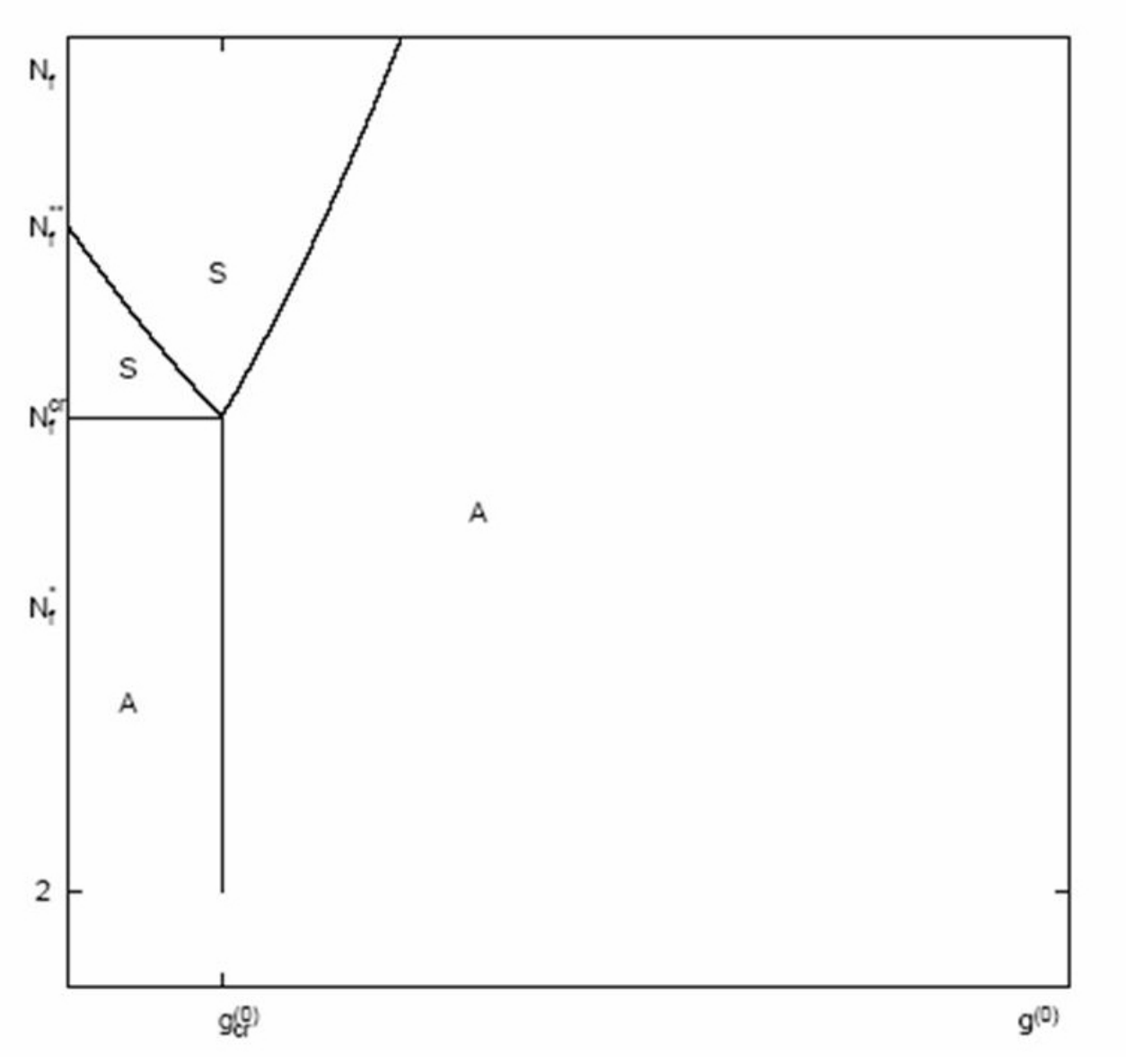}
\caption{The diagram of the various phases of an $SU(N_c)$ gauge theory 
encountered through varying the flavour content $N_f$ and the strength of the coupling $g$. 
At $N_f < N_f^c$ are QCD-like theories. At $N_f > N_f^c$ are theories with a conformal
 phase. For fundamental fermions $N_f^\ast = 8.05$ is the second zero of the two loop 
beta function. S and A refer to symmetric and asymmetric, respectively.}
\label{fig:manyphases}
\end{center}
\end{figure}

Locating an IRFP and revealing a truly conformal behaviour is however an extremely delicate 
task for a lattice simulation. The strategy we have in mind can well be depicted by the 
scenario of \cite{Miransky} and reproduced in Figure~(\ref{fig:manyphases}).   
In this scenario the dynamics of chiral symmetry breaking should  
uplift the lower end of the conformal window $N_f^c$ to values larger than $N_f^\ast$, the 
second zero of the two loop beta function.
At a given $N_f>N_f^c$, and increasing the coupling from $g=0$, one crosses the conformal 
critical line going from a symmetric (S) -- as referred to chiral symmetry --
 asymptotically free phase 
to a symmetric non-asymptotically free phase. At larger couplings a transition to a 
non-symmetric (A) phase will always occur on the lattice, and goes under the name of bulk 
phase transition. The two symmetric phases on the sides of the conformal critical line
 will differ in their short distance behaviour.   

The existence of a bulk phase transition at stronger couplings and the absence of any
thermal transition between a symmetric and a non-symmetric phase will uniquely establish
that the theory is not in the hadronic phase of QCD. Nevertheless, this is not sufficient 
condition to point at the conformal nature of the theory at weaker couplings. Two caveats 
are important in this respect. 
First, non zero fermion masses tend to wash out the infrared fixed point. 
However, sufficiently light fermion masses can be proved to be useful in the vicinity of
 the IRFP: the dependence of the chiral condensate upon light fermion masses should carry a 
measure of the anomalous dimensions at a given lattice coupling. 
Second, the symmetric region on the right of the conformal critical line in 
Figure~(\ref{fig:manyphases}) is not asymptotically free. Thus, an ``improvement'' of 
lattice fermion actions based on the existence of asymptotic 
freedom will loose its reason to be.  

\section{Reflecting on preliminary results}

We discuss some preliminary results obtained with $N_f =12$ flavours in the fundamental 
representation. The set up of the simulations is the same as the one used for the study of 
eight flavours in \cite{PLB_8flavours}.
We use the Asqtad staggered fermion action, with one-loop Symanzik improved and tadpole 
improved gauge action. The rational hybrid monte carlo (RHMC) \cite{Clark:2006wq} is run 
with three pseudofermions. Simulations are performed with the use of the MILC code 
\cite{MILCwebsite} with minor additions.

The drawback of staggered fermions can in this case be the presence of taste symmetry 
breaking. A sizable taste breaking amounts to effectively diminish the active number of 
flavours w.r.t. the nominal one. This effect seems to be drastically reduced in  
recent studies of the QCD spectrum and thermodynamics performed by the MILC collaboration 
\cite{Bernard:2006nj} with improved actions. 
However, an accurate measure of this effect in our context is certainly worth and it is 
ongoing; the study includes the comparison of actions with different type and degree 
of improvement. It has to be added that different quantities will in principle be 
differently affected by taste breaking effects. Our study of phase transition observables
 complemented with the study of the eigenvalue distribution in \cite{Hollands} should 
hopefully shed light on these questions.  
\begin{figure}
\begin{center}
\includegraphics[width=.4\textwidth]{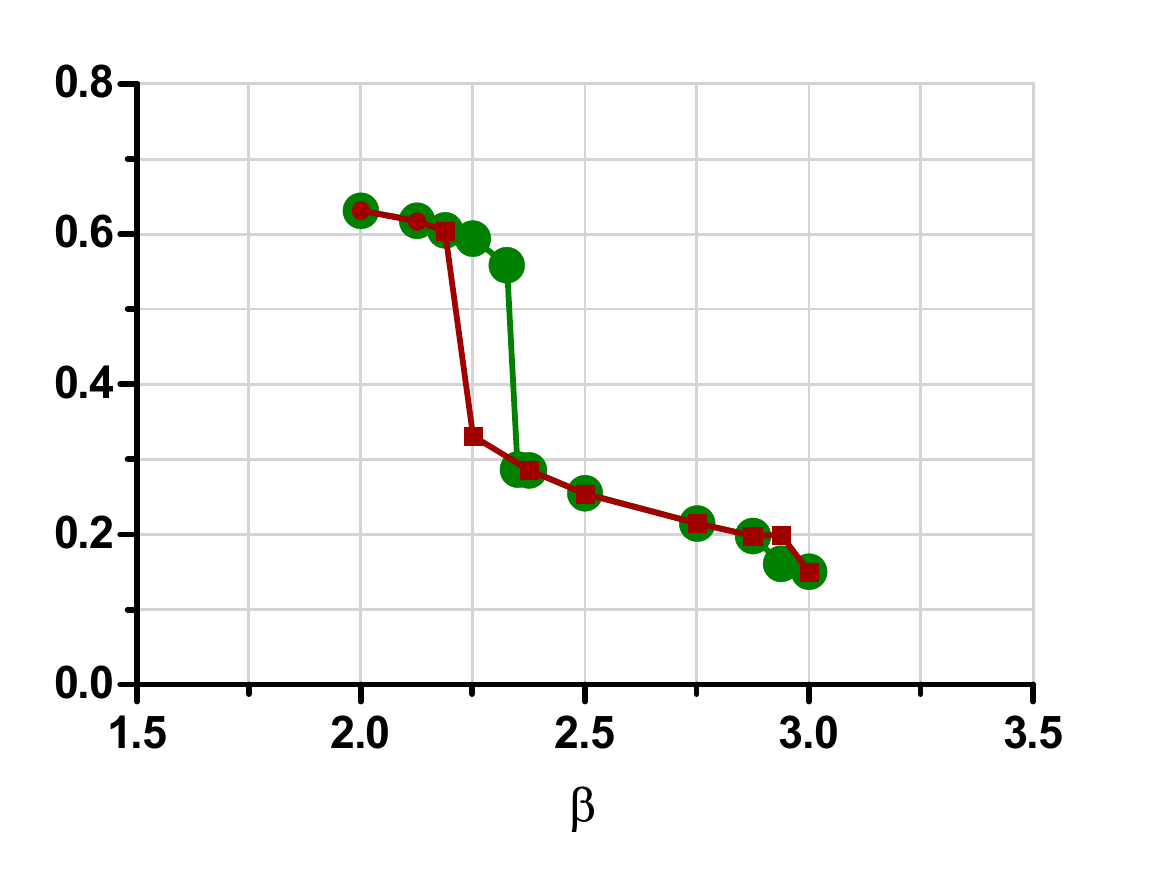}
\caption{The chiral condensate undergoes a transition as a function of the lattice coupling 
$\beta$, for two values of the temporal extent $N_t=8$ (red) and $16$ (green). 
We have suggestive indications that the shift at 
$N_t=8$ might be due to lattice artifacts and not to a thermal nature of the transition.}
\label{fig:bulk}
\end{center}
\end{figure}

We do observe a sharp transition of the chiral condensate at around $\beta = 2.3$ on a 
relatively large lattice with temporal extent $N_t=16$ and spatial volume $V=12^3$. 
This transition might be identified with the bulk transition in the phase diagram of 
Figure~(\ref{fig:manyphases}), if $N_f=12$ turns out to be  already inside the conformal 
window. 

The strategy for excluding the thermal nature of the above mentioned transition is the one 
of excluding a $N_t$ dependence of its location. Though it is important to notice 
that the case of $N_f=12$ might be especially delicate: if $N_f=12$ happens to be close to 
but outside the conformal window, the chiral restoration thermal phase transition will 
happen at almost zero temperature. 
We thus performed simulations with halved temporal extent, on $12^3\times 8$ lattices. 
To our surprise, and as shown in Figure (\ref{fig:bulk}), the transition at 
$N_t=8$ has considerably shifted towards lower couplings -- thus not contradicting a thermal 
nature of the transition -- though remaining equally sharp. 
It must also be said that preliminary results at lower values of $N_t$ would be strongly in 
conflict with the thermal nature of the transition.

Notice that if the transition is to be identified with the bulk transition at strong
coupling in Figure~(\ref{fig:manyphases}), the theory we are describing would be
 {\em not} asymptotically 
free; a lattice action improved on the base of asymptotic freedom might be affected by 
systematic errors which have nothing to do with the physics at hand. We are currently
 investigating this issue, which might be at the origin of the shift observed at $N_t=8$.
Other sources of systematic effects, physical fermion mass dependence, finite spatial volume 
dependence are not expected to explain the observed shift. 

\begin{figure}
\begin{center}
\includegraphics[width=.4\textwidth]{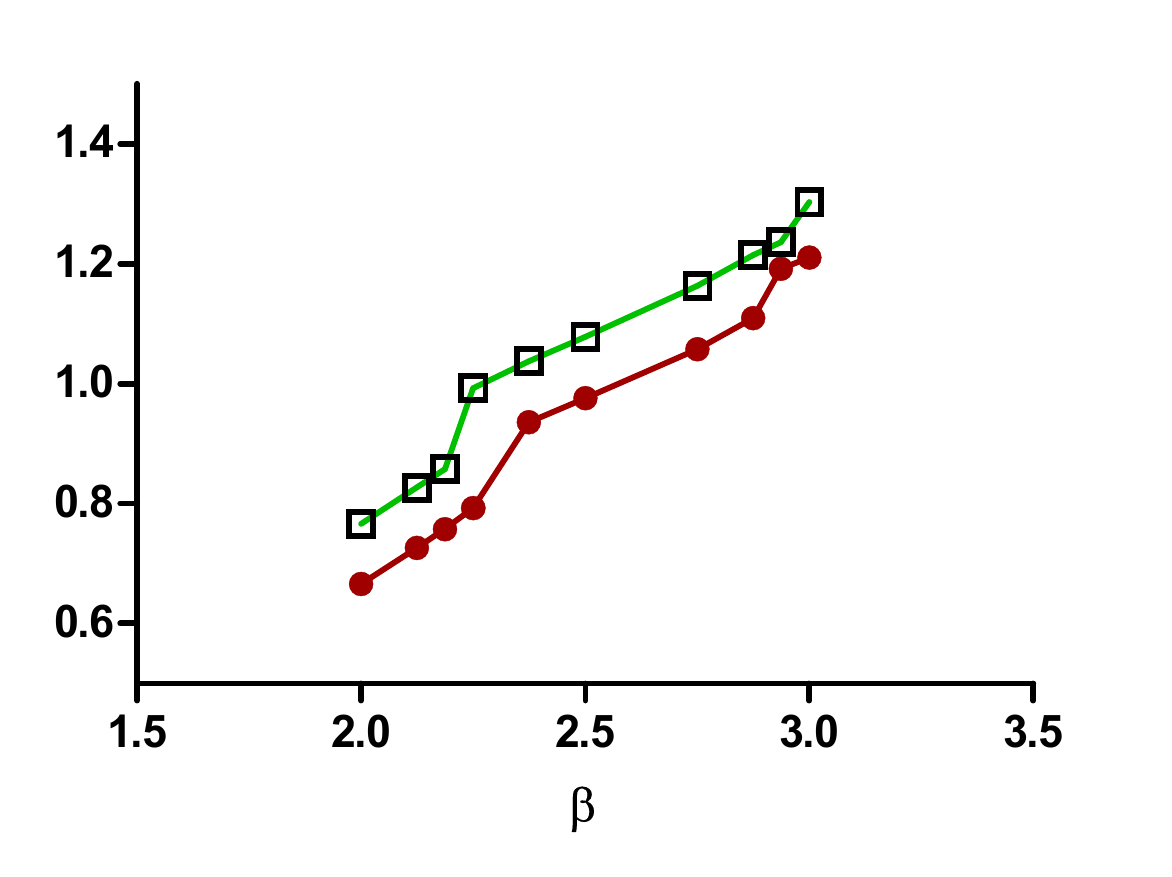}
\caption{The plaquette as a function of the lattice coupling $\beta$. The first jump is
 associated to the jump in the chiral condensate. A tiny jump is also evident at a larger
 value of
 $\beta$.}
\label{fig:plaquette}
\end{center}
\end{figure}

\begin{figure}
\begin{center}
\includegraphics[width=.8\textwidth]{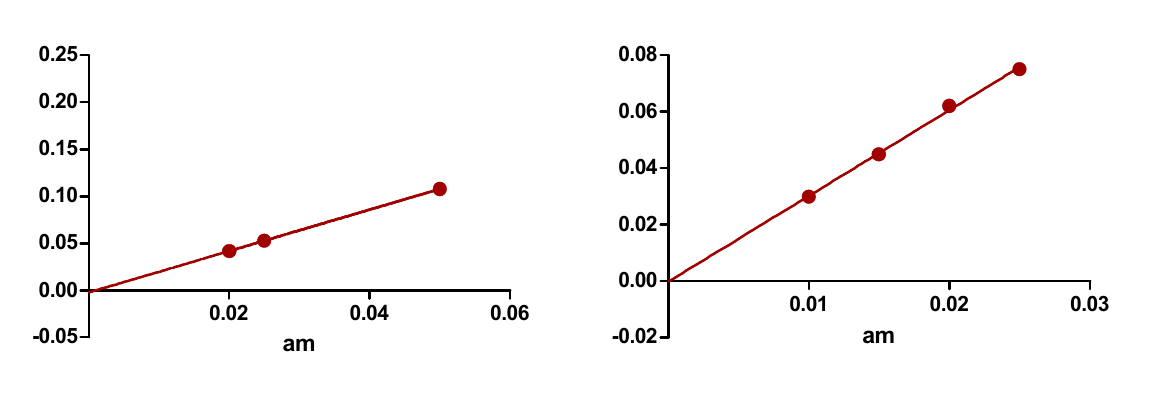}
\caption{Chiral extrapolation of the chiral condensate (at finite lattice spacing!) at 
$\beta = 2.75$ ({\bf left}) and $\beta = 3$ ({\bf right}) .}
\label{fig:chiralb_all}
\end{center}
\end{figure}

We have collected some additional signals which might support the picture of a bulk 
phase transition and the fact that the theory is not in the QCD-like phase.
The plaquette, probe of short distance dynamics, is plotted in Figure (\ref{fig:plaquette}) 
for the two extents $N_t=8$ and $16$ and on a wide range of lattice couplings up to 
$\beta =3$. A second jump in the plaquette is visible. 
Performing a rough chiral extrapolation of the chiral condensate 
(caveat: at finite lattice spacing) on both sides of the second jump (at $\beta = 2.75$ and
 $\beta =3.0$) we obtain a vanishing condensate at zero fermion mass on both sides, 
see Figure 
(\ref{fig:chiralb_all}).
This would suggest that the theory is in a symmetric phase on both sides of the second jump.
A critical analysis is in order. The lattice masses used in the extrapolation at 
$\beta = 2.75$ are still rather heavy and might not allow to detect a possible curvature in 
the extrapolation. However, the masses used at $\beta =3.0$ are already rather light.
The final word can be pronounced once the chiral condensate is measured at lighter masses on 
both sides and the chiral extrapolation is performed after taking 
the continuum limit. 

Should the second jump be interpreted as a true transition? If yes, and if we should assume 
that the chiral condensate extrapolates to zero on both sides of the transition, this would 
make the thermal nature of the first transition -- at stronger coupling -- 
rather improbable. 
All signals till now collected and presented here might well be the realization of the 
scenario in 
Figure~(\ref{fig:manyphases}) for a number of flavours larger than $N_f^c$. According to
 this
scenario, the phase transition observed at $\beta \sim 2.3$ would be a bulk transition, 
while the second 
jump should be a residual signal of the presence of an IRFP at finite values of fermion 
masses. 

To confirm this scenario or disprove it, a careful completion of simulations at a few 
lighter fermion masses, at least one different value of $N_t$, possibly larger than $16$, 
and a careful 
comparison of different types of improved actions and non improved ones is required.
 
In such unexplored terrain is extremely important to use the guidance of perturbation theory 
where possible. We are therefore performing simulations with different actions 
-- staggered improved and unimproved, and Wilson -- at $N_f=16$, where perturbation theory 
can 
be trusted and the location of the IRFP can be perturbatively predicted. 

\acknowledgments 

This work was in part based on the MILC collaboration's public lattice gauge theory code. 
Simulations are performed on the Astron/RUG Stella IBM BG/L in Groningen, and on 
the IBM Power6 Huygens at SARA (Amsterdam).

\end{document}